\documentclass{aastex702}
\usepackage{amssymb}
\usepackage{booktabs}
\newcommand{\epp}{e$^-$\,s$^{-1}$\,pixel$^{-1}$}
\newcommand{\dpd}{$^\circ$\,day$^{-1}$}

\shorttitle{Solar wind in Kepler background photometry}
\shortauthors{Shearer}

\begin{document}

\title{\emph{Kepler} as an accidental heliospheric imager:\\
Thomson-scattered solar-wind structure in four years of background photometry}

\author{Peter M. Shearer}
\affiliation{Scripps Institution of Oceanography, U.C. San Diego, La Jolla, CA 92037, USA}
\email[show]{pshearer@ucsd.edu}

\begin{abstract}
\emph{Kepler} recorded several thousand dedicated sky-background pixels per detector output every
29.4~min from 2009 May to 2013 May while staring at a fixed $\sim12^\circ$ field at ecliptic latitude $\approx+65^\circ$. From this
archive we isolate a large-scale layer that measures brightness variations
relative to an 8-day running median. This layer contains fluctuations of 0.02--0.1~\epp\ ($\sim10^{-17}$ of the mean
solar surface brightness) that sweep across the field in a day or less. Four lines of evidence show that these are
sunlight Thomson-scattered by solar-wind electrons. (1) The sweeps propagate away from the Sun (84\% within $45^\circ$ of the anti-solar direction;
25\% expected by chance), and their direction follows the Sun through each quarter and
across all 16 spacecraft rolls. (2) The field brightness correlates with the near-Earth (OMNI) proton density within
a day of zero lag ($r=0.19$, $p<1/150$), but not with the energetic-proton flux. (3) Time--elongation maps show fronts
moving outward at a median 15\dpd, as forward models of radially moving density structures predict. (4) The amplitude
is about one third of a parameter-free Thomson-scattering prediction and varies by a factor of $\approx3$ over the
year, as predicted. Half of the scattered light comes from within $\approx0.4$--0.7~AU of the spacecraft. Fronts
are strongest in slow wind. \emph{Kepler}'s background pixels thus provide a four-year,
high-cadence, out-of-ecliptic record of solar-wind structure near 1~AU. Standard \emph{Kepler} processing removes
these signals, changing stellar light curves by $\lesssim1$~ppm.
\end{abstract}


\section{Introduction}

Free electrons in the solar wind scatter sunlight, so every line of sight through the heliosphere carries a faint
Thomson-scattered glow proportional to the electron column density weighted by the scattering geometry
\citep[e.g.,][]{Billings1966,Vourlidas2006,HowardTappin2009}. Far from the Sun this signal is $10^{-13}$ of the
solar surface brightness or less and lies hidden beneath zodiacal light and starlight that are
hundreds to thousands of times brighter, so its detection is primarily a problem of background stability.
The first detections came from the zodiacal-light photometers on the \emph{Helios} probes, which were built to map the
brightness and polarization of the interplanetary dust cloud: in searching those data for short-term variations,
\citet{Richter1982} found brightenings caused by interplanetary plasma clouds, and \citet{Jackson1985} used them to
follow coronal mass ejections (CMEs) into interplanetary space. Dedicated heliospheric imagers have since mapped CMEs, corotating
interaction regions (CIRs) and small-scale density structure out to and beyond 1~AU
\citep[e.g.,][]{Eyles2003,Jackson2004,Eyles2009,Howard2013}. Time--elongation maps (``J-maps'') from these
instruments reveal outward-moving fronts whose slopes give their speeds and directions
\citep[e.g.,][]{Sheeley2008,Davies2009,Rouillard2008}, and deep image processing shows structure in the solar wind
down to small scales at 1~AU \citep{DeForest2011}. Interplanetary scintillation tomography provides complementary
three-dimensional reconstructions \citep[e.g.,][]{Jackson1998}.

Like the \emph{Helios} photometers, the \emph{Kepler} mission \citep{Borucki2010,Koch2010} was designed for something quite different: precise
photometry of $\sim1.6\times10^5$ stars in a fixed $\sim115$~deg$^2$ field in Cygnus and Lyra. To estimate the sky
beneath its targets, \emph{Kepler} also downlinked several thousand dedicated background pixels per CCD output at
every 29.4-min long cadence \citep{Jenkins2010}. These pixels, archived but rarely used directly for science, sample the same
patch of sky at fixed positions for four years with excellent photometric stability. In a companion paper
(P. Shearer 2026, submitted to ApJ) we use them to detect the ion tail of comet C/2009~K5 (McNaught) and its disconnection during
the passage of a CME. Here we show that the same data contain a pervasive, mission-long signal of
Thomson-scattered solar wind.

\begin{figure}[p]
\centering
\includegraphics{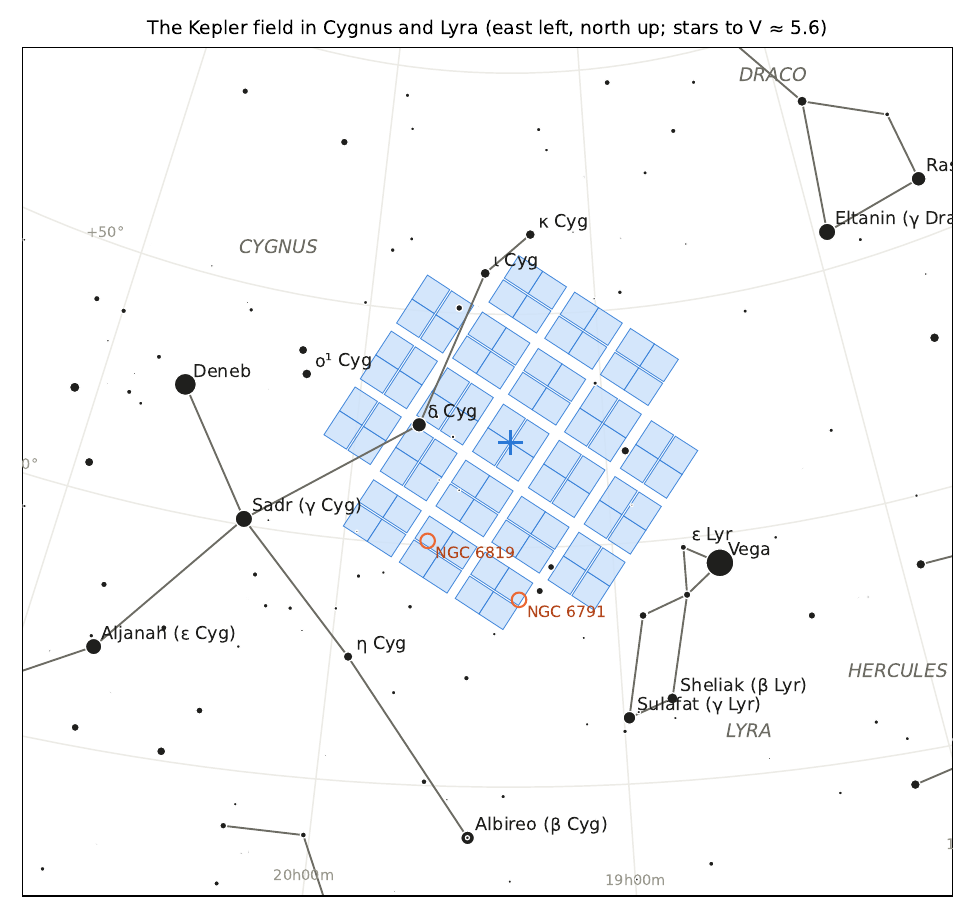}
\caption{The \emph{Kepler} field in Cygnus and Lyra, at ecliptic latitude $\approx+65^\circ$. Over each orbit its
elongation from the Sun varies between $\approx60^\circ$ and $\approx120^\circ$.}
\label{fig:geom}
\end{figure}

\emph{Kepler}'s viewing geometry is unusual for a heliospheric imager. The field lies at ecliptic latitude
$\approx+65^\circ$, and over each 372.5-day orbit its elongation (angular distance from the Sun) swings between
$\approx60^\circ$ and $\approx120^\circ$ (Figure~\ref{fig:geom}). \emph{Kepler} drifted from $\sim0.03$ to
$\sim0.5$~AU behind the Earth during the mission, so the Sun as seen from \emph{Kepler} departs from its
geocentric direction by up to $26^\circ$. Its lines of sight therefore probe solar wind well above the ecliptic
near 1~AU, at elongations beyond the fields of the \emph{STEREO} heliospheric imagers; among dedicated instruments
only the all-sky Solar Mass Ejection Imager (\emph{SMEI}, 2003--2011) routinely covered these directions. The
\emph{Kepler} record (2009--2013) spans the deep minimum between solar cycles 23 and 24 and the rise to cycle-24
maximum.

\section{Data and reduction}
\label{sec:data}

\subsection{Background pixels}
\emph{Kepler}'s background pixels comprise 4200 pixels per CCD channel, arranged as 1050 groups of $2\times2$
pixels on a $\approx2'$ grid avoiding stars \citep{Jenkins2010,VanCleve2016,VanCleve2016b}. We downloaded all 1460 quarterly background files (84 channels in Quarters 0--4, 80 channels thereafter,
after the failure of module~3 in 2010 January) from the Mikulski Archive for Space Telescopes, and reduced each
channel-quarter to a compact record: cadence times, quality flags, the per-cadence coefficients of a robust
two-dimensional surface fit, and float16 residuals for each pixel group (12~GB in total, from $\approx560$~GB of
archival files). The sky level is typically 140--280~\epp. Zodiacal light provides most of it and all of its
seasonal variation \citep{Leinert1998}; unresolved Galactic starlight adds a constant part that increases toward the Galactic plane, by
$\approx6$~\epp\ per degree of decreasing Galactic latitude $b$ across the field ($b=6^\circ$--$21^\circ$), as measured
from the background levels of the individual CCD outputs \citep[cf.][]{Leinert1998}.

\subsection{Surface fits}
For each channel and cadence we fit a robust cubic surface in focal-plane coordinates to the pixel groups. The
field-wide signal studied here is carried by these surfaces, not by the residuals: within a single channel
($\approx1.1^\circ$ across) the solar-wind brightness is nearly uniform.

\subsection{Large-scale layer and masks}
For each channel we evaluate the surface on the group positions, with its cubic terms smoothed over 1~day (a cubic
refitted at every cadence can absorb real structure within the channel), and subtract its own 8-day running median.
This ``large-scale layer'' retains changes on time scales from 30~min to a few days. We mask the first
3~days of each quarter and 2~days after each data gap (thermal settling after rolls and downlinks); cadences
affected by solar energetic particles, identified where the cadence-to-cadence scatter exceeds 1.25 times the
channel median (widened by $\pm6$~hours); and isolated electronic glitches (jumps $>10\sigma$ in a channel's
spread). The layer is binned onto a $30'$ sky grid aligned with the focal plane, at 3-h and 30-min resolution, and
expressed in \epp.  Sky positions of the pixel groups come from cubic maps, one per channel, that take the flux-weighted
centroids of stars to their catalogued positions.

\emph{Minimal version and robustness.} None of these details is critical. Repeating the main analyses with a single
number per channel and cadence, the fitted background level at the channel center minus its 8-day running median,
reproduces every result (Table~\ref{tab:robust}). Switching off the masks one at a time, or all except the pipeline
quality flags, also leaves the conclusions unchanged; only the energetic-particle mask noticeably improves the
precision of the direction tracking.

\begin{table}[t]
\centering
\caption{Main results from the full large-scale layer and from the per-channel central level alone.}
\label{tab:robust}
\resizebox{\textwidth}{!}{\begin{tabular}{lccc}
\toprule
Result & large-scale layer & channel level & channel level, no masks$^a$\\
\midrule
Directions within $45^\circ$ of anti-solar$^b$ & 84\% & 91\% & 92\%\\
Mean offset from anti-solar & $-12^\circ$ & $-13^\circ$ & $-13^\circ$\\
Slope of direction vs.\ anti-solar PA & $1.01\pm0.03$ & $1.01\pm0.09$ & $0.89\pm0.18$\\
Windows with fronts / outward / median speed & 60\% / 88\% / 15\dpd & 61\% / 89\% / 15\dpd & 60\% / 92\% / 15\dpd\\
Annual modulation, exponent on $G$ & $1.19\pm0.15$ & $1.08\pm0.12$ & $1.22\pm0.14$\\
Peak correlation with OMNI density & 0.19 & 0.14 & 0.13\\
\bottomrule
\end{tabular}}
\\[2pt]{\footnotesize $^a$ Only pipeline quality flags applied. $^b$ Windows with gradient significance $>10$.}
\end{table} 

\subsection{Sun direction as seen from \emph{Kepler}}
Because \emph{Kepler} trails the Earth, all geometric quantities use the direction of the Sun as seen from the
spacecraft, taken from daily JPL Horizons ephemerides with \emph{Kepler} as observer \citep{Giorgini1996}. An
independent solution from the ephemerides of $\approx460$ comets computed for the same observer agrees to $0.005^\circ$.

\section{Results}

\subsection{Sweeps propagate away from the Sun}
\label{sec:direction}
\begin{figure}[p]
\centering
\begin{interactive}{animation}{sw_fig2_anim.mp4}
\includegraphics[width=\textwidth]{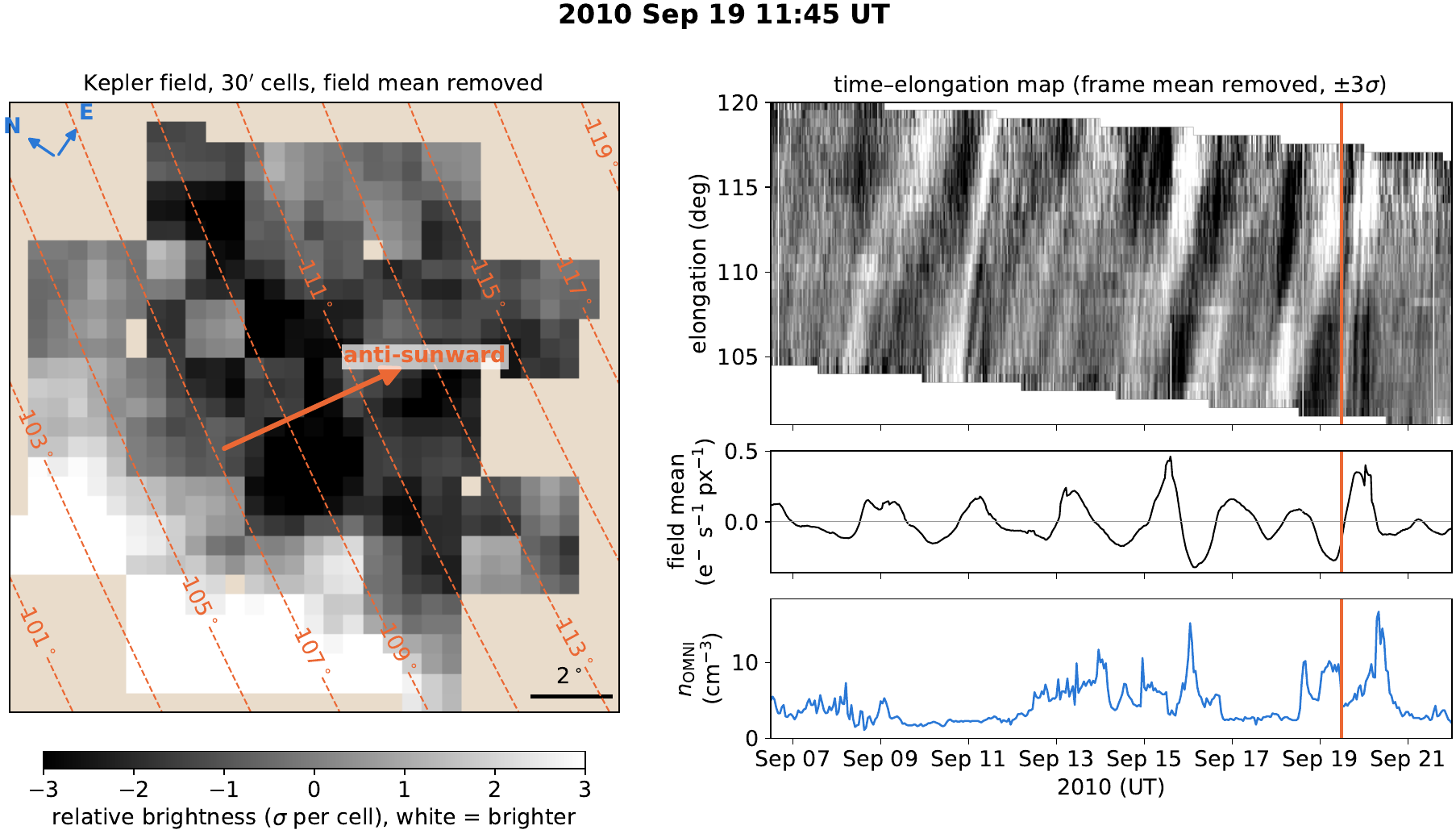}
\end{interactive}
\caption{One frame (2010 September 19 11:45 UT) from the animation of the large-scale layer in Quarter~6. Left: the
\emph{Kepler} field in $30'$ cells (axes aligned with the CCD array; blue arrows: north and east). Each cell is
band-passed (2-day running mean removed) and smoothed over 3~h; the field mean is then removed and each cell is
divided by its own robust scatter (white = brighter). Dashed lines: elongation from the Sun as seen from
\emph{Kepler}; arrow: anti-sunward direction. Right, top to bottom: the time--elongation map for this interval (as in
Figure~\ref{fig:jmap}), the field-mean brightness (band-passed and smoothed like the cells), and the OMNI proton
density; the orange line marks the frame time. In this frame a bright front has entered the field at the lowest
elongations, just before the field-mean brightening of September 19--20. An animation of this figure is available.
It runs from 2010 September 6 12:00 to September 22 00:00 UT in 30-min steps (744 frames) and lasts 31~s; fronts
appear as bright and dark bands parallel to the elongation contours that move anti-sunward across the field.}
\label{fig:movie}
\end{figure}

Movies of the large-scale layer (Figure~\ref{fig:movie} and its animation) show brightness changes that sweep across the field in about a day or less. To measure their
direction we divide the mission into overlapping 3-day windows (step 1~day). In each window we band-pass every $30'$ cell
(removing a 2-day running mean), cross-correlate it with the field-mean series to obtain its delay, and fit the
delays with a plane; the gradient gives the propagation direction and apparent speed, and its ratio to its
uncertainty gives a significance.

\begin{figure}[p]
\centering
\includegraphics{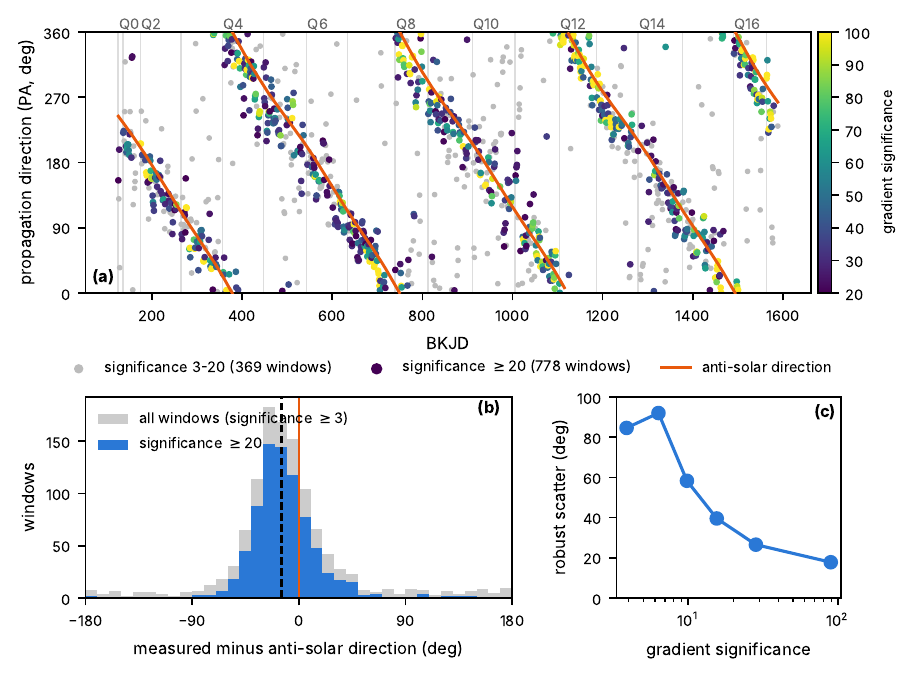}
\caption{Propagation direction of large-scale brightness sweeps. (a) Direction in 3-day windows (step 1~day); colored
points have arrival-time gradient significance $\ge20$, gray points 3--20. The orange line is the anti-solar position
angle at the field center as seen from \emph{Kepler}; vertical lines mark the quarterly rolls. (b) Distribution of the
offset from anti-solar for all windows (gray) and for significance $\ge20$ (blue; dashed line: median). (c) Robust
scatter of the direction as a function of gradient significance.}
\label{fig:direction}
\end{figure}

Of 1006 windows with significance $>10$, 84\% propagate within $45^\circ$ of the anti-solar direction, compared
with 25\% expected for random directions; for the 597 most significant windows (significance $>30$) the fraction
is 93\% (Figure~\ref{fig:direction}). Because the windows overlap, the number of independent measurements is about
one third of these totals, but the preference for the anti-solar direction remains overwhelming. The anti-solar position angle at the field rotates by $\sim80^\circ$ within
each quarter as the Sun moves, while the spacecraft attitude is fixed. The measured direction follows the Sun, with a
weighted mean slope of $1.01\pm0.03$ over 17 quarters. The quarterly $90^\circ$ rolls produce no discontinuities in the sweep directions. A pattern
fixed to the focal plane or spacecraft structure (for example rolling-band electronic noise, heater cycling or
focus changes) would keep a constant direction within a quarter and jump at each roll. The propagation is
therefore tied to the Sun, not to the instrument.

Individual directions scatter widely when the arrival-time gradient is weak, that is, when the field brightens almost
simultaneously. The robust scatter falls from $\approx90^\circ$ for gradient significance below 8 to $21^\circ$ for the
778 windows with significance above 20, 90\% of which lie within $45^\circ$ of anti-solar
(Figure~\ref{fig:direction}b,c).
The directions are not exactly anti-solar. Relative to the anti-solar direction seen from \emph{Kepler}, the average
sweep direction is rotated by a nearly constant $-12^\circ$ (mean; quarterly means $-5^\circ$ to $-22^\circ$) toward
the direction of solar rotation. This can be seen in the offset of the points in Figure~\ref{fig:direction}a from the
anti-solar line and in the histograms of the angular deviations in Figure~\ref{fig:direction}b. We return to this
offset in Section~\ref{sec:cir}.

\subsection{Correlation with near-Earth solar-wind density}
\label{sec:omni}
\begin{figure}[p]
\centering
\includegraphics{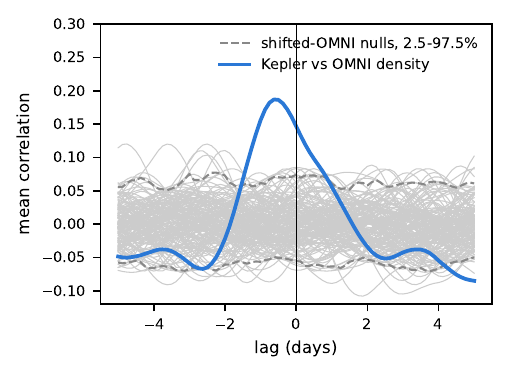}
\caption{Correlation between the \emph{Kepler} field-mean brightness and the OMNI proton density as a function of
lag, averaged over quarters (blue), compared with 150 nulls using time-shifted OMNI data (gray; dashed lines are the
2.5th and 97.5th percentiles). Negative lags mean that \emph{Kepler} leads.}
\label{fig:omni}
\end{figure}

If the signal is Thomson scattering by solar-wind electrons, it should relate to the solar-wind density. We compare
the field-mean large-scale brightness (8-day running median removed, 1-day smoothing) with the hourly OMNI proton
density \citep{King2005}, processed identically, for lags of $\pm10$~days. Averaged over quarters, the correlation
peaks at $r=0.19$ at a lag of $-0.6$~day, with \emph{Kepler} leading. At zero lag $r=0.15$, and the correlation is positive in 14 of 17
quarters. To assess significance we repeat the calculation with the OMNI series shifted by 150 random offsets of
40--500~days, avoiding multiples of the 27.3-day solar rotation to exclude recurrent structure; the 99th
percentile of the null peak correlation within $\pm2$~days is 0.09, and no null reaches the observed value
(Figure~\ref{fig:omni}).

The observed lag should not be over-interpreted. Corotating structure passes the plasma that \emph{Kepler} sees before or
after it reaches the Earth, depending on the heliocentric longitude of the plasma. Integrating the Thomson weight
along \emph{Kepler}'s lines of sight, we predict that this lead varies from quarter to quarter between $-0.4$ and
$+3$~days, growing as \emph{Kepler} drifts behind the Earth. The peak lags of individual quarters (from the
per-channel series of Table~\ref{tab:robust}) scatter by about $\pm1$~day, are uncorrelated with this prediction
(Spearman $\rho=-0.01$), and are near zero in both 2009--2010 and 2012--2013. The correlation therefore shows that
the \emph{Kepler} signal follows the near-Earth density, but the lag does not locate the plasma or measure how much
of the structure corotates.

Energetic particles accompany CME-driven shocks and stream interfaces and could in principle raise the background
through residual particle hits. The correlation is specific to density (Table~\ref{tab:omni}): excluding all
intervals of elevated $>10$ or $>60$~MeV proton flux (10\% of the data) leaves it essentially unchanged (0.18),
while the proton fluxes themselves correlate no better than the null. Solar-wind speed correlates more weakly and
peaks 2~days later, as expected because compressed slow wind precedes the arrival of fast streams
\citep[e.g.,][]{GoslingPizzo1999,Richardson2018}.  The correlation is modest because the two quantities sample different plasma. OMNI measures the density at a single point upstream of the Earth. \emph{Kepler}'s lines of sight integrate over $\sim1$~AU, rise steeply out of the
ecliptic, and start from a point 0.03--0.5~AU behind the Earth.

\begin{table}[t]
\centering
\caption{Correlation of the \emph{Kepler} field-mean brightness with OMNI quantities (mean over quarters).}
\label{tab:omni}
\begin{tabular}{lcc}
\toprule
OMNI quantity & peak $r$ & lag of peak (days)\\
\midrule
Proton density & 0.19 & $-0.5$ to $-0.6$\\
Proton density, particle-quiet intervals only & 0.18 & $-0.5$\\
Solar-wind speed & 0.11 & $+2.0$\\
Proton flux $>10$~MeV (log) & 0.09 & $-3.0$\\
Proton flux $>60$~MeV (log) & 0.08 & $-1.8$\\
Null (shifted OMNI), 99th percentile & 0.09 & --\\
\bottomrule
\end{tabular}
\end{table}

\subsection{Time--elongation maps and front speeds}
\label{sec:jmaps}
\begin{figure}[p]
\centering
\includegraphics{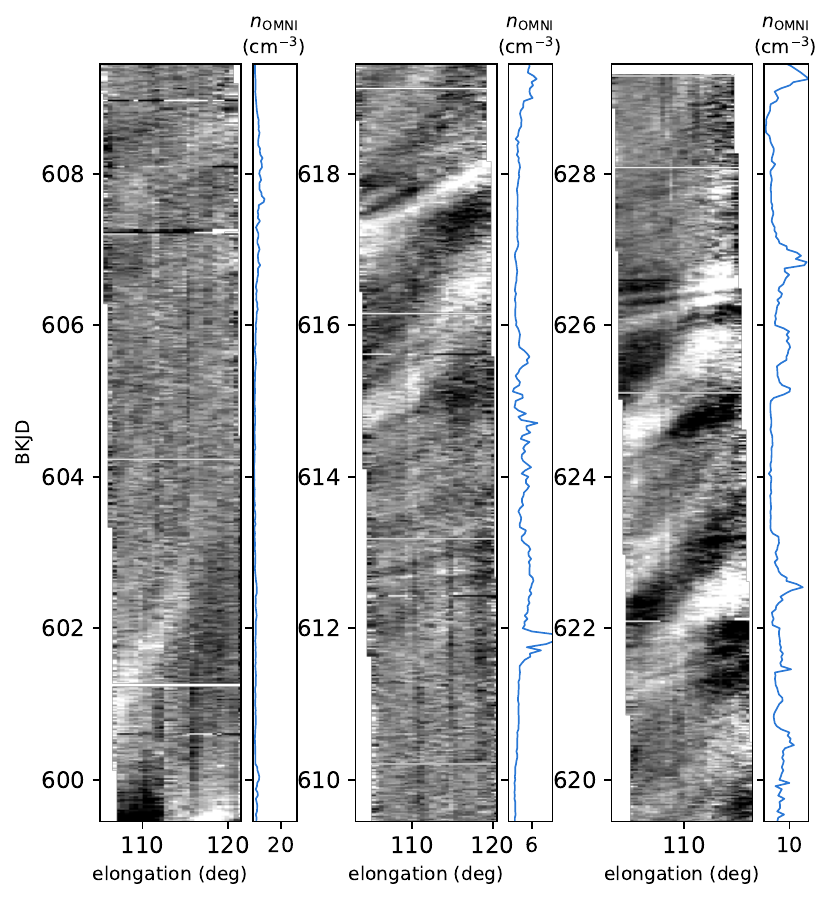}
\caption{Time--elongation maps for Quarter~6 (2010 September), 30-min resolution, frame mean removed and normalized
by the noise in each elongation column ($\pm3\sigma$; white = brighter). Fronts slanting up and to the right move
away from the Sun. Beside each column: OMNI proton density (blue), and near-Earth ICME disturbances and intervals
(orange; \citealt{Richardson2010}).}
\label{fig:jmap}
\end{figure}

Since the sweeps propagate along the Sun--anti-Sun direction, much of their information can be collapsed into
time--elongation maps. At each cadence we compute the elongation of every $30'$ cell, average the cells in
$0.5^\circ$ elongation bins, and stack the profiles in time (Figure~\ref{fig:jmap}). After removing the frame mean,
and normalizing each elongation column by its robust noise within the quarter, outward-moving structures appear as
fronts slanting toward larger elongation with time, as in J-maps from dedicated heliospheric imagers
\citep[e.g.,][]{Sheeley2008,Davies2009}. Fronts are common during active intervals and absent during quiet ones,
and are strongest in the quarters closest to the Sun.

We measure front speeds by slant-stacking the frame-mean-removed 30-min maps in 2-day windows: each elongation
column is shifted by $(\varepsilon-\bar\varepsilon)/\omega$ for trial angular speeds $\omega$ between $\pm3$ and
$\pm1000$\dpd, and the speed that maximizes the variance of the stacked signal is retained when it exceeds the median
over trials by a factor of three. Of 2401 windows, 1455 meet this criterion; 88\% of these are outward, with median
speed $15$\dpd\ (16th--84th percentiles 10--24\dpd), nearly independent of elongation (15, 15 and 13\dpd\ at
field-center elongations below $80^\circ$, $80^\circ$--$100^\circ$ and above $100^\circ$). Repeating the analysis with the elongation labels of the columns shuffled at random, which destroys any coherent
propagation, gives fronts in only 19\% of windows, of which 49\% are outward, with slower speeds (median 7\dpd);
the observed speeds and the outward preference are therefore not produced by the method. Removing the frame mean
is essential: delays
measured from the raw maps are biased toward zero by any component that brightens the whole field at once, and give
speeds three to four times higher.

\subsection{Thomson-scattering model}
\label{sec:model}
We model the signal with a point Sun, an electron density $n_e=n_1(r/1~\mathrm{AU})^{-2}$, and Thomson scattering
with phase function $1+\cos^2\chi$, where $\chi$ is the scattering angle \citep[e.g.,][]{Billings1966,HowardTappin2009}.
The brightness contributed per unit length along a line of sight at distance $D$ from \emph{Kepler} is
\begin{equation}
\frac{dB}{dD} = F_\odot\,\Omega_{\rm px}\; n_e(r)\,\frac{r_e^2}{2}\,(1+\cos^2\chi)\left(\frac{1~\mathrm{AU}}{r}\right)^2,
\end{equation}
where $F_\odot$ is the solar flux at 1~AU in \emph{Kepler} units ($1.74\times10^5$~e$^-$\,s$^{-1}$ at $K_p=12$ and
$K_p(\odot)=-26.9$), $\Omega_{\rm px}$ is the solid angle of a $3.98''$ pixel, and $r_e$ is the classical electron
radius.

\begin{figure}[p]
\centering
\includegraphics{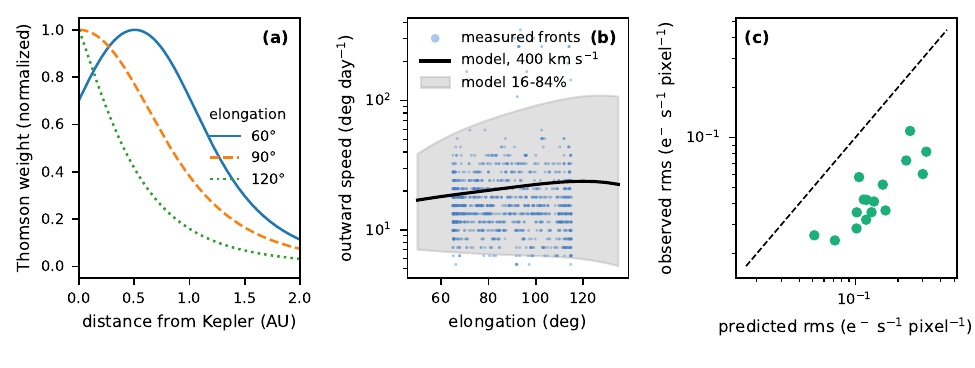}
\caption{Thomson-scattering model. (a) Weighting along the line of sight at three elongations. (b) Measured
outward front speeds (points) and the Thomson-weighted median prediction for radial motion at 400~km\,s$^{-1}$
(line; shading 16th--84th percentiles); see Table~\ref{tab:synth} for forward models analyzed like the data.
(c) Observed versus predicted rms brightness fluctuation per quarter; the dashed line is equality.}
\label{fig:model}
\end{figure}

\emph{Where the signal comes from.} The weighting function (Figure~\ref{fig:model}a) is broad. At $60^\circ$
elongation it peaks $\approx0.5$~AU from the spacecraft (median 0.71~AU); at $90^\circ$ and $120^\circ$ it is
highest next to the spacecraft, with medians of 0.55 and 0.39~AU. The median heliocentric distance of the
scattering plasma is $\approx1.0$~AU at $65^\circ$ elongation and $\approx1.2$~AU at $115^\circ$. Because the lines
of sight point $\approx65^\circ$ out of the ecliptic, its median ecliptic latitude is high: $\approx40^\circ$ at
$65^\circ$ elongation, $26^\circ$ at $90^\circ$ and $17^\circ$ at $115^\circ$.

\emph{Brightness.} With $n_1=1$~cm$^{-3}$ the model gives 0.10--0.25~\epp\ for field-center elongations of $113^\circ$--$66^\circ$.
For the observed OMNI densities the total Thomson brightness is 0.4--1.5~\epp, under 1\% of the zodiacal
background. The observed robust rms of the band-passed field mean is 0.024--0.11~\epp\ (3--8\% of the predicted
total). This is a factor of $\approx3$ (median 0.33) below a prediction that assumes the whole line of sight
fluctuates coherently with the OMNI density. A reduction of this size is expected if the fluctuations come from
structures $\sim0.1$~AU across, many of which are averaged along $\sim1$~AU of line of sight. We have not measured
the structure size independently, however, so the amplitude is a consistency check rather than a confirmation of
the model. The quarter-to-quarter amplitudes track the prediction (Figure~\ref{fig:model}c; Spearman $\rho=0.81$,
$p\approx10^{-4}$, 16 quarters). This reflects the steep elongation dependence of the scattered light discussed in
Section~\ref{sec:annual}, and is not an independent test.

\emph{Speeds.} A Thomson-weighted median for radially moving structures at the OMNI speed predicts 18--22\dpd,
increasing slowly with elongation; the measured speeds are 0.77 times this on average (16th--84th percentiles
0.45--1.40). A median along the line of sight is not, however, what the slant-stack measures. We therefore built a
forward model in which synthetic density structures are placed in three dimensions around \emph{Kepler}'s actual
orbit and lines of sight, their Thomson brightness is computed every 3~h, and the resulting maps are analyzed with
exactly the same procedure as the data (Table~\ref{tab:synth}; 12 realizations per entry). Radially moving shells
(CME-like structures launched at random within $\pm30^\circ$ heliolatitude) at 400~km\,s$^{-1}$ reproduce the observed
speeds at all three elongations (medians 12--17\dpd); at 650~km\,s$^{-1}$ they are too fast (25--28\dpd) except at the
largest elongations. Corotating Parker-spiral density walls give similar speeds at small and intermediate
elongations but slower fronts at $\approx112^\circ$ (8.5\dpd), and their apparent speed is set by solar rotation
rather than by the wind speed. Both models give outward fronts in 70--85\% of windows, fewer than the observed 88\%.
The front speeds therefore favor radially moving structures at typical wind speeds but do not exclude a corotating
contribution (Section~\ref{sec:cir}).

\begin{table}[t]
\centering
\caption{Front speeds from the slant-stack: data and forward models processed identically. Model entries are medians
over 12 realizations (four 30-day intervals in different years $\times$ three random seeds), with the 16th--84th
percentile range across realizations; trial speeds lie on a logarithmic grid (8.5, 9.9, 11.5, 13.3, 15.4, 17.9, 20.8,
24.1, 28.0\dpd, \ldots), so values cluster at grid steps. Observed values are for field-center elongations below
$80^\circ$, $80^\circ$--$100^\circ$ and above $100^\circ$.}
\label{tab:synth}
\resizebox{\textwidth}{!}{\begin{tabular}{lccc}
\toprule
& \multicolumn{3}{c}{median outward speed (\dpd) at field-center elongation}\\
& $\approx67^\circ$ & $\approx91^\circ$ & $\approx112^\circ$\\
\midrule
Observed & 15 & 15 & 13\\
Radial shells, $V=400$~km\,s$^{-1}$ & 13 (13--21) & 17 (13--18) & 12 (9--13)\\
Radial shells, $V=650$~km\,s$^{-1}$ & 28 (24--28) & 25 (24--28) & 13 (11--15)\\
Corotating spiral walls, $V=400$~km\,s$^{-1}$ & 13 (13--15) & 13 (11--13) & 8.5 (8.5--8.5)\\
Thomson-weighted median (radial, OMNI speed, analytic) & 18 & 20 & 22\\
\bottomrule
\end{tabular}}
\end{table}

\subsection{Annual modulation and the radial decline of density fluctuations}
\label{sec:annual}
\begin{figure}[p]
\centering
\includegraphics{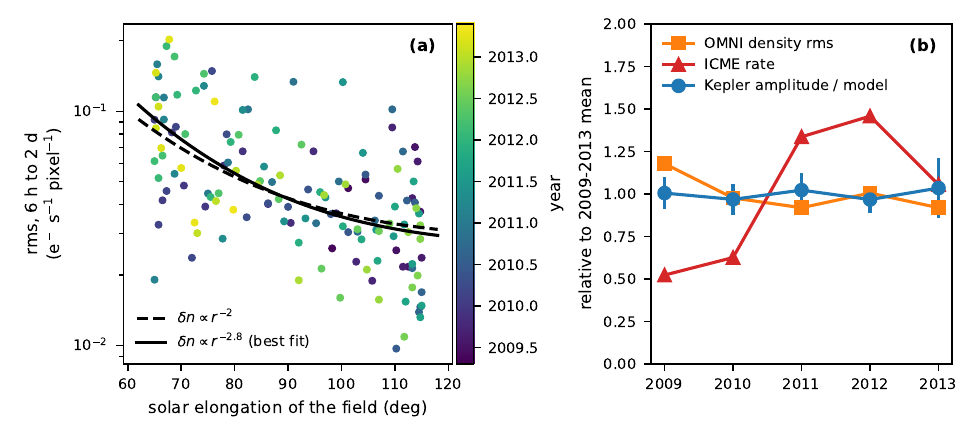}
\caption{Annual modulation. (a) Rms of 6-h to 2-day variations of the field-mean brightness in ten-day segments
(white noise removed), versus the solar elongation of the \emph{Kepler} field; color shows the date. Dashed: Thomson
model for fluctuations proportional to the mean density ($\delta n\propto r^{-2}$); solid: best fit,
$\delta n\propto r^{-2.8}$. (b) Year by year, each relative to its 2009--2013 mean: \emph{Kepler} amplitude divided by
the model brightness per unit density (blue; standard errors), rms near-Earth density variation in the same band
(OMNI, orange), and the rate of near-Earth interplanetary CMEs during the \emph{Kepler} mission (red;
\citealt{Richardson2010}).}
\label{fig:annual}
\end{figure}

The \emph{Kepler} field lies at ecliptic latitude $65^\circ$, so its solar elongation changes from $65^\circ$ (late
January) to $115^\circ$ (late July) as \emph{Kepler} orbits the Sun. At small elongation the lines of sight pass
inside 1~AU through the peak of the scattering weight; beyond $90^\circ$ they look outward from the spacecraft and
sample plasma farther from the Sun (median heliocentric distance 0.98~AU at $65^\circ$, 1.24~AU at $115^\circ$;
Figure~\ref{fig:model}a). A given fractional density fluctuation should therefore appear 2.6 times brighter in
winter than in summer (0.257 versus 0.099~\epp\ per cm$^{-3}$ at 1~AU). We measured the rms of 6-h to 2-day
variations of the field mean in 139 ten-day segments, removing the white-noise contribution estimated from
cadence-to-cadence differences (Figure~\ref{fig:annual}a). The amplitude falls by a factor of 2.8, from 0.085~\epp\
at elongations of $60^\circ$--$75^\circ$ to 0.030~\epp\ at $108^\circ$--$120^\circ$, so the seasonal cycle is as large
as the variation between successive ten-day intervals. Writing the amplitude as $\propto G^a$, where $G(\varepsilon)$
is the model brightness per unit density, and allowing a separate normalization for each year, gives
$a=1.19\pm0.15$ for the field mean and $1.39\pm0.14$ for the elongation gradient, which carries the fronts. Using only
the change of elongation within each quarter gives $a=1.5\pm0.3$.

The seasonal contrast is not an instrumental effect proportional to the sky brightness. The zodiacal sky is
brighter at small elongation, but only by a factor of 1.6 (median 229 versus 143~\epp), and the white noise changes by
only a factor of 1.3, whereas the fluctuations change by 2.8 and their elongation gradient, which carries the
propagating fronts, by 3.5.

\emph{Radial decline of the fluctuations (tentative).} Taken at face value, an exponent above unity means that the
fluctuations weaken with distance from the Sun faster than the mean density. If the fluctuating density scales as
$\delta n\propto r^{-k}$, the seasonal contrast gives $k=2.8$ ($1\sigma$ range 2.2--3.5) for fluctuations that are
coherent along the line of sight. If the signal is instead the sum of many independent structures along the line of
sight, the seasonal contrast for $k=2$ is smaller ($a=0.7$--0.8 for structures 0.1--0.4~AU across, after averaging
over the $\sim12^\circ$ field), and a larger $k$ would be needed. We regard this inference as tentative. It rests on a
modest range of heliocentric distance ($\approx1.0$--1.2~AU) seen from a single viewpoint, through a spherically
symmetric model. The winter lines of sight also sample higher latitudes (median ecliptic latitude $\approx40^\circ$,
versus $\approx17^\circ$ in summer), so any latitudinal structure of the wind enters the seasonal contrast. Around
solar minimum, when the dense, structured slow wind is concentrated near the ecliptic, this would weaken the winter
signal, but the latitude structure changed as solar activity rose during the mission. Since compressions at stream
interfaces are expected to grow with distance from the Sun \citep[e.g.,][]{GoslingPizzo1999}, a decline steeper than
$r^{-2}$ would be notable, and it should be tested against heliospheric-imager and in-situ measurements of density
structures at different distances \citep[e.g.,][]{Stansby2018}.

\emph{No solar-cycle dependence.} At fixed viewing geometry the yearly mean amplitudes from 2009 to 2013 agree within
their standard errors ($\approx10\%$; Figure~\ref{fig:annual}b), although the rate of near-Earth interplanetary CMEs
\citep{Richardson2010} nearly tripled, from 12.5 to 35 per year, as solar activity rose. The near-Earth density
variability in the same band (OMNI) is also nearly constant (0.8--1.0~cm$^{-3}$ rms). On these time scales the
\emph{Kepler} signal therefore appears to be set mainly by the ambient solar wind, with CMEs producing conspicuous
individual events (Section~\ref{sec:icme}) but little of the overall level of fluctuation. Two aspects of the
processing work against a CME contribution, however: the energetic-particle mask removes intervals around the
largest, shock-driving CMEs, and the 8-day baseline removes slower variations. The conclusion applies to 6-h to 2-day
variations outside energetic-particle events.

\subsection{Slow wind, corotating structure and the direction offset}
\label{sec:cir}
Splitting the windows by the OMNI speed at the time (Table~\ref{tab:wind}) shows that fronts are more frequent
(66\% vs.\ 40\% of windows) and stronger (median rms 0.081 vs.\ 0.067~\epp\ in windows with fronts) in slow ($<380$~km\,s$^{-1}$) than in
fast ($>480$~km\,s$^{-1}$) wind, at all elongations. Dense, variable slow wind therefore dominates the signal, as
expected for scattering that scales with density and for the known abundance of density structure in slow wind
\citep[e.g.,][]{Sheeley1997,Viall2015}. Fronts during fast wind are somewhat slower (11.5 vs.\ 15\dpd), closer to the
corotating model.

The $-12^\circ$ offset of the propagation direction from anti-solar (Section~\ref{sec:direction}) points the same
way. For plasma moving radially, the apparent motion seen from \emph{Kepler} is along the sky projection of
$V\hat{\mathbf r}-\mathbf v_K$, where $\mathbf v_K$ ($\approx30$~km\,s$^{-1}$) is the spacecraft's orbital velocity.
This aberration rotates the expected direction by $+4.5^\circ$, opposite to the observed offset, so the data lie
$-16^\circ$ from the expected radial direction. A corotating density wall along the Parker spiral advances along
its normal, which at 1~AU points $\approx45^\circ$ from radial in the direction of solar rotation
\citep{Parker1958}; such structures are routinely tracked in heliospheric images \citep[e.g.,][]{Rouillard2008,Plotnikov2016}.
Projected onto \emph{Kepler}'s field, this rotates the expected direction by $-41^\circ$
(16th--84th percentiles $-48^\circ$ to $-34^\circ$), with little dependence on distance along the line of sight. A
mixture of radial and corotating structure could therefore explain the offset; a simple linear mixture would imply
that roughly 40\% of the signal comes from corotating structure. This fraction is illustrative only. The projected
motion of a CME-like shell launched away from the line of sight is not exactly anti-solar either, the mixture has not
been tested with a forward model of the two-dimensional sky motion, and the OMNI lag gives no independent support
(Section~\ref{sec:omni}). The offset is similar in slow, intermediate and fast wind ($-15^\circ$ to $-16^\circ$), so
if it is due to corotating structure, that structure contributes throughout and not only at stream interfaces.

\begin{table}[t]
\centering
\caption{Front statistics by near-Earth solar-wind speed (OMNI median within $\pm1$~day of each 2-day window). A window has
fronts when its slant-stack contrast exceeds 3; the median speed is over outward fronts, and the median rms (of the
high-passed time--elongation map) is over the windows with fronts.}
\label{tab:wind}
\resizebox{\textwidth}{!}{\begin{tabular}{lccccc}
\toprule
Wind speed & windows & with fronts & outward & median speed (\dpd) & median rms (\epp)\\
\midrule
$<380$~km\,s$^{-1}$ & 1189 & 66\% & 87\% & 15.4 & 0.081\\
380--480~km\,s$^{-1}$ & 899 & 60\% & 90\% & 13.3 & 0.076\\
$>480$~km\,s$^{-1}$ & 313 & 40\% & 88\% & 11.5 & 0.067\\
\bottomrule
\end{tabular}}
\end{table}

\subsection{Interplanetary CMEs}
\label{sec:icme}
\begin{figure}[p]
\centering
\includegraphics{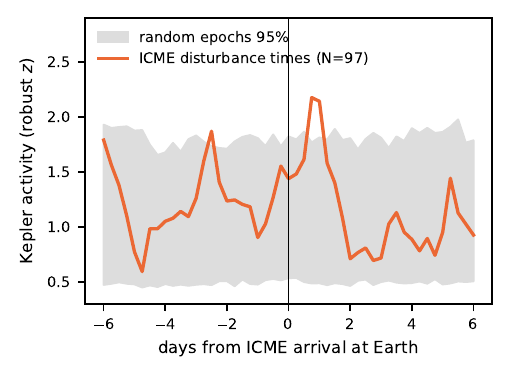}
\caption{Superposed-epoch analysis of \emph{Kepler} large-scale activity around the arrival at the Earth of 97
interplanetary CMEs \citep{Richardson2010}, compared with random epochs (gray, 95\% range).}
\label{fig:icme}
\end{figure}

Some intervals rich in fronts coincide with known interplanetary CMEs \citep{Richardson2010}, and a CME launched on
2010 April 13 is a candidate trigger for the tail disconnection of comet C/2009~K5 (companion paper). Such
coincidences are easy to find after the fact, however, so we test the association statistically. A
superposed-epoch analysis of all 97 near-Earth ICMEs with \emph{Kepler} coverage shows a peak in \emph{Kepler}
activity 0.75~days after the disturbance reaches the Earth (Figure~\ref{fig:icme}). This peak exceeds the 95\%
range of random epochs at that lag but is not significant once all lags are considered ($p=0.15$). That is
unsurprising: many Earth-directed ICMEs miss \emph{Kepler}'s out-of-ecliptic lines of sight, and CMEs crossing those
lines of sight need not reach the Earth. Converting the angular speeds of single events into radial speeds requires
their distance along the line of sight, which a single viewpoint does not provide; we therefore report event speeds
in angular units (6--15\dpd\ medians) and leave their interpretation to multi-viewpoint comparisons.

\section{Discussion}

\paragraph{Instrumental alternatives.}
Brief diffuse brightenings of the focal plane in \emph{Kepler} data (``Argabrightenings''; \citealt{VanCleve2016b}) last one or two cadences and have
no preferred direction. Effects fixed to the spacecraft are excluded by the tracking of the Sun direction through
the rolls, and Sun-driven stray light would not propagate outward, correlate with solar-wind density or prefer slow
wind.

\paragraph{Implications for \emph{Kepler} stellar photometry.}
Most users of \emph{Kepler} data work with the PDCSAP light curves, so it is important to ask whether this
previously unrecognized signal biases them. Two pipeline steps act against it. First, the background is estimated
at every cadence by fitting a low-order two-dimensional polynomial to the background pixels of each CCD output and
subtracting it from the target apertures \citep{Jenkins2010,Twicken2010}. Because the Thomson-scattered signal varies
smoothly over the $\approx1.1^\circ$ extent of an output, this fit absorbs nearly all of it. Second, PDC removes
systematics shared by many stars \citep{Smith2012,Stumpe2012}.
We tested the end product directly. For each quarter we took 20\,000 randomly chosen stars with $K_p=14.5$--16 (the
faint stars, for which a background error matters most), formed the median of their fractional PDCSAP deviations at
each cadence, applied the same band-pass as in Section~\ref{sec:omni}, and regressed it on the large-scale
background signal. If the background were not subtracted at all, the expected response is
$N_{\rm ap}/F_\star\approx1100$~ppm per \epp\ for a 12-pixel aperture. The measured response is
$-0.05\%\pm0.05\%$ of this when averaged over 17 quarters, with no quarter exceeding $0.8\%$ in absolute value.
For the observed fluctuation amplitudes (0.02--0.08~\epp\ rms), any residual is $\lesssim1$~ppm for stars of
$K_p=15$--16 and proportionally smaller for brighter stars. That is far below the photometric noise on any time
scale. The undetected residual also implies that the pipeline removes more than 99\% of the signal. This test
measures background subtraction and PDC co-trending together and does not separate them. The
Thomson-scattered solar wind therefore has no practical effect on \emph{Kepler} light curves, including searches
for transits and stellar variability. Diffuse sources with structure on scales of arcminutes, such as comet comae
and ion tails, are not captured by the background fit and can leave brief, localized signals in the light curves of
faint stars; they are discussed in the companion paper.

\paragraph{What \emph{Kepler} adds.}
Dedicated instruments image Thomson-scattered solar wind in different ways. The \emph{STEREO} heliospheric imagers
orbit the Sun near 1~AU, like \emph{Kepler}, and so avoid scattered light from the Earth and Moon, but they look
toward the Sun at elongations of $\approx4^\circ$--$89^\circ$, behind deep baffles that reject direct sunlight
\citep{Howard2008,Eyles2009}. \emph{SMEI} imaged nearly the whole sky, including the \emph{Kepler} field, from low
Earth orbit from 2003 to 2011, at a resolution of about $1^\circ$ and with a background complicated by aurorae,
energetic particles and Earth and Moon light \citep{Eyles2003,Jackson2004}; \emph{PUNCH}, also in low Earth orbit,
images out to $\approx45^\circ$ from the Sun \citep{DeForest2026}. \emph{Kepler}'s background pixels add a
30-min cadence, four-year record of a fixed $\sim12^\circ$ field whose lines of sight rise steeply out of the ecliptic
near 1~AU, at elongations of $65^\circ$--$115^\circ$, observed from a deep-space orbit with high photometric
stability. In the units used for heliospheric imagers (S10, the brightness of one tenth-magnitude star per square
degree; 1~S10 $\approx1.34$~\epp\ for \emph{Kepler}), the field-mean fluctuations are 0.015--0.075~S10, and a direct
comparison of sensitivity with \emph{SMEI} for the same field would be worthwhile. The record overlaps the all-sky \emph{SMEI} data (2003--2011) and the \emph{STEREO} heliospheric imagers
\citep{Howard2008,Eyles2009}, enabling multi-viewpoint studies of individual structures, and three-dimensional
reconstructions from interplanetary scintillation and \emph{SMEI} tomography \citep[e.g.,][]{Jackson1998,Jackson2011}
could be integrated along \emph{Kepler}'s lines of sight to predict its brightness directly. The same approach should
apply to the K2 mission (2014--2018), whose fields lie in the ecliptic and whose later campaigns overlap renewed
\emph{STEREO-A} coverage, and in principle to \emph{TESS} (2018--present). A first look at the pipeline background
estimates of \emph{TESS} 2-min targets, however, is dominated by scattered light from the Earth and Moon, which
\emph{TESS}'s orbit makes far stronger than for \emph{Kepler}; solar wind detection from \emph{TESS} will likely require the full-frame images and an explicit scattered-light model.

\paragraph{Limitations.}
The sky sampling is sparse ($2\times2$-pixel groups on a $2'$ grid) and the field is only $\sim12^\circ$ across, so
individual structures are seen over a small range of elongation. Distances along the line of sight are not measured
directly. The 8-day baseline and the cleaning remove much of the signal on time scales longer than a few days.
The correlation analysis assumes that near-Earth density represents the plasma \emph{Kepler} sees.

\section{Conclusions}
\begin{enumerate}
\item \emph{Kepler}'s background pixels record field-wide brightness fluctuations of 0.02--0.1~\epp\ that propagate
away from the Sun and follow its direction continuously through four years and 16 spacecraft rolls.
\item The fluctuations correlate with the near-Earth solar-wind density, independently of energetic particles, and
are strongest in slow wind.
\item Their front speeds agree with forward models of radially moving density structures within $\sim1$~AU of the
spacecraft, and their amplitudes are consistent with Thomson scattering by structures $\sim0.1$~AU across.
\item The amplitude changes by a factor of $\approx3$ over the year as the field's elongation changes, as Thomson
scattering predicts, but not with the solar cycle. The seasonal contrast tentatively suggests that density
fluctuations decline with heliocentric distance faster than the mean density, and the nearly constant level from
2009 to 2013 suggests that the ambient wind, rather than CMEs, sets the overall level of 6-h to 2-day fluctuations.
\item The standard \emph{Kepler} pipeline processing (background subtraction and co-trending) removes the solar wind signal: it changes PDCSAP light
curves of faint stars by $\lesssim1$~ppm, so \emph{Kepler} stellar photometry is unaffected.
\item A systematic rotation of the propagation direction toward the direction of solar rotation may indicate a
contribution from corotating, Parker-spiral structure.
\end{enumerate}

\begin{acknowledgments}
This paper includes data collected by the \emph{Kepler} mission and obtained from the MAST data archive at the Space
Telescope Science Institute. We used the OMNI data from NASA/GSFC's Space Physics Data Facility, the
Richardson \& Cane near-Earth ICME catalog, and JPL Horizons. Analysis and narration assisted by Claude (Anthropic).
\end{acknowledgments}

\facilities{Kepler}

\software{NumPy \citep{Harris2020}, Matplotlib \citep{Hunter2007}}

\section*{Data and code availability}
The \emph{Kepler} background-pixel files are available from MAST, OMNI data from the NASA Space Physics Data Facility,
the ICME catalog from \citet{Richardson2010} and its online updates, and ephemerides from JPL Horizons. The reduced
background archive and the code used for the analyses, figures and animation are available on Zenodo: \dataset[doi:10.5281/zenodo.24182076]{https://doi.org/10.5281/zenodo.24182076}. All the \emph{Kepler} long-cadence data used in this paper can be found in MAST:
\dataset[doi:10.17909/T9488N]{https://doi.org/10.17909/T9488N}.

\end{document}